\documentclass[aip,prb,twocolumn, unsortedaddress,superscriptaddress]{revtex4}

\usepackage{graphicx}
\usepackage[T1]{fontenc}
\usepackage{textcomp}

\newcommand{\CO}{$^{12}$CO }
\newcommand{\COn}{$^{12}$CO}

\newcommand{\COt}{$^{13}$CO }
\newcommand{\COtn}{$^{13}$CO}

\begin{document}

%Title of paper
\title{A traveling wave decelerator for neutral polar molecules}
\author{Samuel A. Meek}
\affiliation{Fritz-Haber-Institut der Max-Planck-Gesellschaft, Faradayweg 4-6, 14195 Berlin, Germany}
\author{Maxwell F. Parsons}
\affiliation{Fritz-Haber-Institut der Max-Planck-Gesellschaft, Faradayweg 4-6, 14195 Berlin, Germany}
\affiliation{Physics Department, Harvard University, Cambridge, Massachusetts 02138, USA}
\affiliation{Ecole Polytechnique F\'ed\'erale de Lausanne, Institut des Sciences et Ing\'enierie Chimiques, 1015 Lausanne, Switzerland}
\author{Georg Heyne}
\author{Viktor Platschkowski}
\author{Henrik Haak}
\author{Gerard Meijer}
\email[]{meijer@fhi-berlin.mpg.de}
\affiliation{Fritz-Haber-Institut der Max-Planck-Gesellschaft, Faradayweg 4-6, 14195 Berlin, Germany}
\author{Andreas Osterwalder}
\email[]{andreas.osterwalder@epfl.ch}
\affiliation{Ecole Polytechnique F\'ed\'erale de Lausanne, Institut des Sciences et Ing\'enierie Chimiques, 1015 Lausanne, Switzerland}

\date{\today}

\begin{abstract}
Recently, a decelerator for neutral polar molecules has been presented that operates on the basis of macroscopic, three-dimensional, traveling electrostatic traps (Osterwalder et al., Phys. Rev. A {\bf 81}, 051401 (2010)).
In the present paper, a complete description of this decelerator is given, with emphasis on the electronics and the mechanical design.
Experimental results showing the transverse velocity distributions of guided molecules are shown and compared to trajectory simulations.
An assessment of non-adiabatic losses is made by comparing the deceleration signals from \COt with those from \CO and with simulated signals.
\end{abstract}

%\pacs{37.10.Mn,37.10.Pq,37.20.+j}
%\keywords{}

\maketitle

\section{Introduction}
Interest in the production of cold molecules stems from a variety of research areas, most notably collision dynamics, the study of long-range interactions, quantum computation, and high-resolution spectroscopy (for a recent overview see the individual chapters of reference 1).% \cite{Krems:2009wc}).

Two principal approaches can be distinguished in the production of cold molecules: either the molecules are assembled from ultracold atoms using electric fields, magnetic fields, or lasers, or one starts from warm or fast samples and brings them to low average translational temperatures by controlling the velocity.\cite{Krems:2009wc,vandeMeerakker:2008ii}
Variants of the latter approach include using collisions with a cold buffer gas\cite{Doyle:1995p274} or between molecules in crossed-beams,\cite{Elioff:2003p586} by mechanical means such as rotating nozzles,\cite{Gupta:2001wq,Strebel:2010dg} or through the controlled interaction with electric,\cite{Bethlem:1999ir,Bethlem:2000kg,Bethlem:2003da} magnetic,\cite{Vanhaecke:2007ge,Narevicius:2007jw,LavertOfir:2010wj} or optical fields.\cite{Fulton:2006iq}

One simple way to use electric fields to passively control a molecular velocity distribution is by filtering out fast molecules using the competition between an electrostatic force and a centrifugal force in a bent guide.\cite{Rangwala:2003cy,Sommer:2009cc,Bertsche:2010vr}
This does not change the velocity of the individual molecules but merely extracts the slow ones from a thermal distribution.
Active control of the forward velocity of molecules with electric fields was first demonstrated in 1999,\cite{Bethlem:2000kg,Bethlem:1999ir} rapidly followed by the first trapping of neutral molecules.\cite{Bethlem:2000ti,Bethlem:2002cy}
Further developments of this technology have enabled the deceleration of large molecules in high-field-seeking states,\cite{Wohlfart:2008ia} the trapping of molecules in static and ac-fields,\cite{Bethlem:2002cy,vanVeldhoven:2005gd} and the storage of molecules in a synchrotron.\cite{Heiner:2009ie,Zieger:2010p5336}

For most of the studies mentioned above, considerable density both in velocity space and in position space are of great importance.
A significant development in Stark deceleration was thus the recognition that such devices should be operated in a higher-order mode in order to suppress the coupling of longitudinal and transverse oscillations and the losses following from them.\cite{vandeMeerakker:2005kz,Scharfenberg:2009ht}
Recently, a new approach has been presented to completely avoid this problem, namely the use of traveling three-dimensional traps to capture and decelerate molecules.\cite{Osterwalder:2010bx}
This approach builds on ideas implemented in a miniaturized Stark decelerator on a chip,\cite{Meek:2009er,Meek:2009dg,Meek:2008ex} and a similar principle has also been applied to the deceleration of Rydberg states\cite{Vliegen:2007jj}.
In contrast to a conventional Stark decelerator, where effective moving traps are obtained by periodic switching of the electric fields between two static configurations, these devices employ continuous modulation of the voltages to produce real, continuously moving three-dimensional traps.
In the chip decelerator, this modulation is chosen to produce a chain of traps above a periodic structure of electrodes, which in this case are stripes on a dielectric substrate.
The modulation frequency determines, for a particular electrode geometry, the velocity of the traps and can be chirped to produce acceleration or deceleration.
In the present decelerator, the stripes of the chip decelerator are replaced with three-dimensional ring electrodes.\cite{Osterwalder:2010bx}
The decelerator is scaled up by a factor of 100 in comparison with the chip decelerator, which results in dimensions comparable to conventional Stark decelerators.
Based on trajectory calculations, the six-dimensional phase space acceptance for guiding of OH molecules at a constant velocity is determined to be a factor of three larger than in the state-of-the-art switched Stark decelerator,\cite{Scharfenberg:2009ht} and during deceleration, the same relative phase space acceptance is maintained for an acceleration which is a factor of three larger.
Furthermore, the molecules are loaded into the traps at high velocity, and the loading of stationary traps is simply bringing the traveling traps to standstill.

The present paper gives a detailed account of the design and construction of this new decelerator and shows experimental data together with the corresponding trajectory simulations to explain the operation principle of the device and the dynamics of individual molecules inside the decelerator.
The paper is structured as follows: section \ref{ch:exp} describes the general principle of operation (\ref{ch:op}), the mechanical design (\ref{ch:mec}), and the required electronics (\ref{ch:el}).
Section \ref{ch:mec} includes the particular approach chosen to assemble the structure.
Section \ref{ch:sim} presents experimental results that characterize the apparatus, and theoretical calculations to support the interpretation of these measurements.
The paper is summarized in section \ref{ch:sum} where also an outlook is given, along with future applications of this design.

\section{Experimental}\label{ch:exp}
The following sections give a general overview on the operation principle of the decelerator and describe in detail how the device is designed and constructed and how the required electronics are built. 

\subsection{General considerations}\label{ch:op}
\begin{figure}
\includegraphics[width=\linewidth]{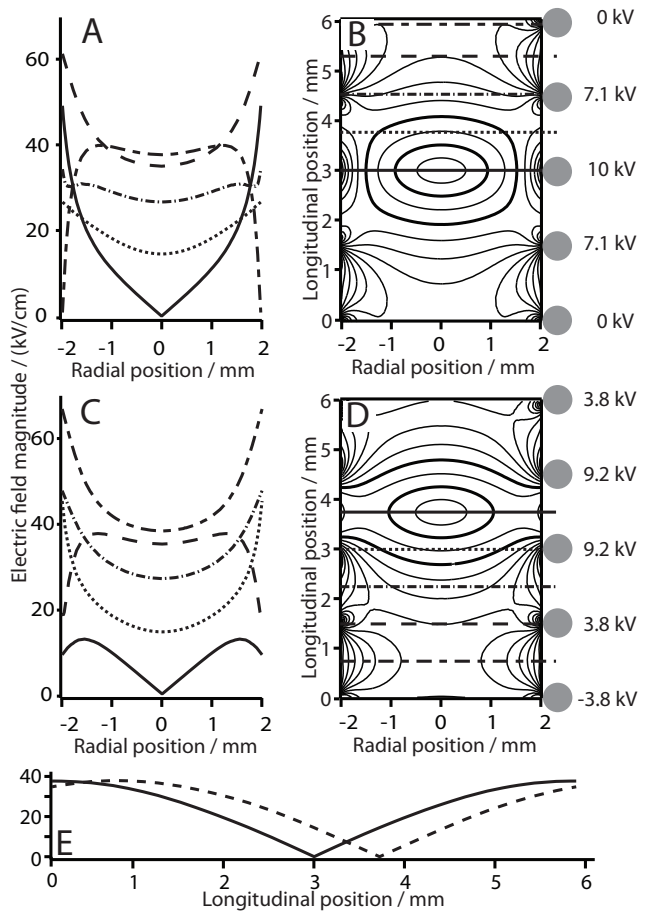}%
\caption{\label{fig:figpot}Calculated electric fields for the trap located in the plane of an electrode (A and B), and between two electrodes (C and D). Panels A and C show the radial dependence of the electric field magnitude, panels B and D show contour plots of the electric field magnitude. Line types in panels A and C correspond to cross sections at the positions indicated by lines of the same type in panels B and D. Contours start at 5 kV/cm around the center of the trap and are spaced by 5 kV/cm. The thick lines represent 10 kV/cm and 20 kV/cm. Gray circles on the right of panels B and D show the positions of the electrodes. Numbers next to the circles indicate the voltages applied to the individual electrodes for both cases.
Panel E shows the longitudinal dependence of the electric field magnitude for the cases from panels B (solid line) and D (dashed line).}
 \end{figure}

Applying a spatial sine-modulated electric potential to a periodic array of ring electrodes produces an electric field distribution on the inside of the array with two on-axis minima in the field magnitude per period (see figure \ref{fig:figpot}).
Due to the cylindrical symmetry, these minima represent three-dimensional traps for polar molecules.

The required potential on the individual electrodes can be expressed as
\begin{equation}
V_n(t)=V_0\sin(-\phi_0(t)+2\pi n/N)
\end{equation}
with the waveform amplitude $V_0$ and the time-dependent phase offset $\phi_0(t)$.
The electrode-dependent shift is determined through the periodicity $N$ of the specific array.
A linear increase of $\phi_0(t)$ with time continuously shifts the traps along the cylinder axis.
The time-dependence of the phase offset can be expressed as
\begin{equation}
\phi_0(t)=2\pi\int_0^t\nu(\tau)d\tau,
\end{equation}
where the frequency $\nu(\tau)$ is the modulation frequency of the voltage on each individual electrode.
Since one oscillation of the waveform moves the trap over one period, the velocity of the traps is given by $L\nu(\tau)$, where $L$ is the period length.
The time-dependence of the frequency leads to acceleration or deceleration of the traps.
After calculating the acceptance for many different geometrical parameters, a suitable combination of these was found with 4 mm inner electrode diameter, 0.6 mm electrode thickness, 12 mm period length, and periodicity 8.
Lower periodicity means fewer points to sample the spatial sine, and thus less smooth electric field distributions.
Higher periodicity improves the electric fields but makes the setup technically more challenging.
The field distribution produced by a larger number of electrodes per period alone is not necessarily better: in optimizing the number of electrodes, it was found that -- while keeping all other parameters the same -- a decelerator with 8 electrodes per period performs better than one with 10, though a decelerator with 6 is significantly worse than either of these.

In order to create sufficiently deep traps moving at the required velocity for this geometry, the electronics must produce waveforms with $V_0\approx$10 kV, with a tunable frequency up to $\approx$25 kHz (corresponding to a velocity of 300 m/s).
The phase of the waveforms between neighboring electrodes is shifted by $\pi/4$.

Contour plots for single traps resulting from $V_0$=10 kV sine potentials used with this design are shown in figures \ref{fig:figpot}B and D for the cases where the trap is positioned in the plane of an electrode (corresponding to $\phi_0=0$) and between two electrodes ($\phi_0=\pi/8$), respectively.
Contours are spaced by 5 kV/cm, starting at 5 kV/cm.
Thick lines show the contours for 10 kV/cm and 20 kV/cm, respectively.
The gray circles to the right of panels B and D show the positions of the electrodes.
Numbers next to the circles show the voltages that are applied to each electrode in the two cases.
Panels A and C show the radial dependence of the electric field magnitude at positions designated by the lines with corresponding line types in panels B and D.
Each line represents the cross section for different distances from the trap minimum.
In both cases, the solid, dotted, dash-dotted, long-dashed, and long-short-dashed lines represent 0 mm, 0.75 mm, 1.5 mm, 2.25 mm, and 3 mm from the trap minimum; 3 mm from the trap minimum is the position of the saddle point toward the next trap.
The on-axis longitudinal potential is shown in panel E for the two cases.
Here, the solid (dashed) line represents the case with $\phi_0=0$ ($\phi_0=\pi/8$).
Panel E shows that, longitudinally, the potential does not change when moving the trap from in-plane to between two electrode planes.
This is also confirmed through the height of the saddle point in panels A and C.
Longitudinally, the traps are almost 40 kV/cm deep when sine waves with amplitudes of 10 kV are applied to the electrodes.
For a linearly Stark-shifted state, the trap has the shape of $|\sin(x)|$, which is nearly linear close to the minimum.
While the shape of the longitudinal potential is very stable and independent of $\phi_0$, the radial potential does undergo some oscillations.
The trap is almost 50 kV/cm deep when located in the plane of an electrode, but $\approx$13 kV/cm when between two electrodes.
For the $a\, ^3\Pi_1$($v = 0$, $J = 1$) state of CO used here this leads to a transverse acceptance of $\approx\pm$22 m/s and $\approx\pm$11 m/s, respectively.
Trajectory simulations on a virtual decelerator with 10 m length demonstrated that the effective transverse acceptance lies in between these values and is close to the one obtained when averaging the radial potential along the guide axis.
Longitudinally, the acceptance for guiding at constant velocity is $\approx\pm$19 m/s.\cite{Osterwalder:2010bx}

\subsection{Mechanical design and construction}\label{ch:mec}
\begin{figure}
\includegraphics[width=\linewidth]{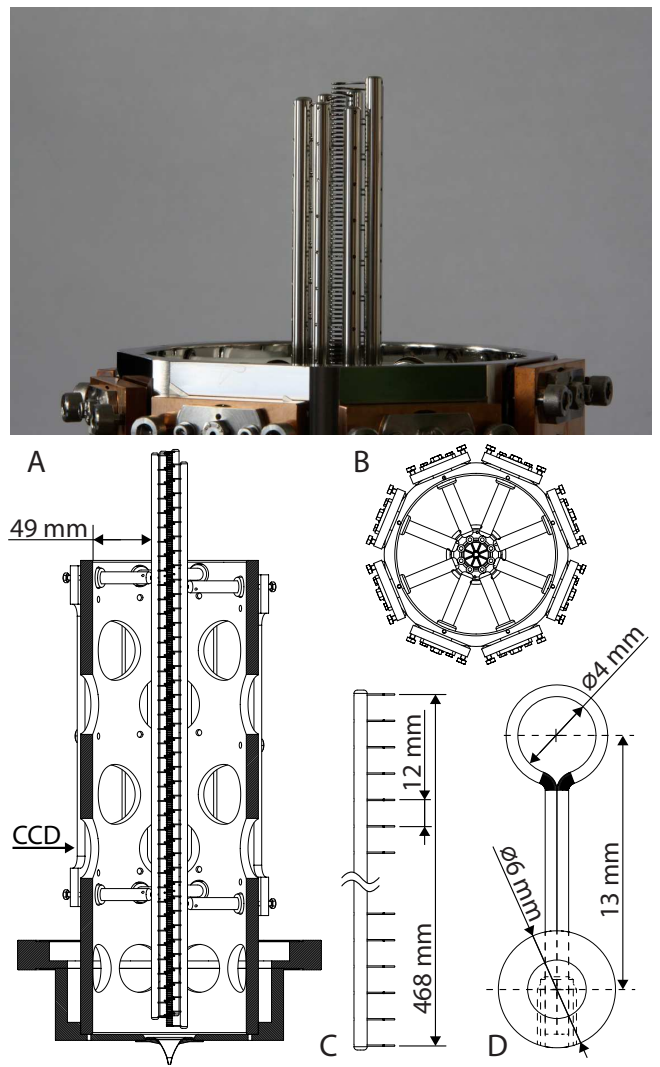}%
\caption{\label{fig:fig1} Upper section: Photograph showing the upper end of the decelerator structure. Bottom section: Mechanical design of the traveling wave decelerator. A) Cross section through the entire decelerator. The arrow labeled ``CCD'' shows the position of the CCD camera used for the phosphorescence imaging. B) Top view of the decelerator and mount structure. C) Single rod containing one set of 40 ring electrodes. D) Top view of a single electrode with cross section through the mounting rod. The active element of the electrode is the circular section at the top in this figure.}
\end{figure}
The upper part of figure \ref{fig:fig1} is a photograph of the top end of the decelerator.
The central part shows the electrodes, visible as horizontal bars, and the electrode mounting rods, as described below.
The lower part of the photograph shows the mounting structure with the main octagonal stainless-steel tube and bronze plates that hold the electrode mounts.
An overview of the decelerator design, including its mount, is shown in figure \ref{fig:fig1}A.
Figures \ref{fig:fig1}B-D show, respectively, a top view of the decelerator, a single electrode mount, and an electrode with a cross section through the mount.
The decelerator is positioned vertically inside a high-vacuum chamber.
Molecules are produced in a pulsed supersonic expansion and enter the decelerator chamber through a skimmer, visible at the bottom of figure \ref{fig:fig1}A, that is placed as close as possible to the beginning of the decelerator.
In the case of the CO experiments presented below, a UV laser crosses the expansion directly before the skimmer to excite the CO molecules to the upper $\Lambda$-doublet component of the metastable $a\, ^3\Pi_1$($v = 0$, $J = 1$) state.
This particular state has a phosphorescence lifetime of 2.63 ms.\cite{Gilijamse:2007wda}

The decelerator is formed from 320 ring electrodes that are mounted on 8 separate 6 mm-diameter stainless steel rods.
It is worth remembering that the number of electrodes here can not be compared to the number of deceleration stages in a traditional Stark decelerator;
the figure of merit is instead the electric field gradient that can be generated, and this determines the length of the decelerator, which in turn defines the number of electrodes.
The 8 rods are each mounted on ceramic posts that are attached via adjustable bronze bars to the main octagonal mounting structure (see figure \ref{fig:fig1}B).
The mechanical point of reference of the entire setup is the center of the base flange that holds the skimmer and supports the decelerator.
This construction, and the vertical orientation of the decelerator, greatly simplifies the alignment by making it independent of the main vacuum chamber. 
A magnification of one of the rods is shown in panel \ref{fig:fig1}C.
The individual electrodes are spaced by 12 mm which defines the period length.
All 8 rods, hand-picked for straightness, are identical and are mounted with a longitudinal offset of 1.5 mm from their nearest neighbors, a value which is given by the required center-to-center distance of neighboring electrodes.
The arrangement of successive electrode rods is counter-clockwise when seen from the skimmer, forming a left-handed helical pattern.

A single electrode is shown in panel \ref{fig:fig1}D.
The active part of these electrodes is the top, circular part with an inner diameter of 4 mm.
The bottom part is the handle via which the electrode is mounted to the main rod.
Each electrode is formed from 0.6 mm diameter Ta wire in a three-step procedure:
\begin{enumerate}
\item The wire is stretched, polished, and cut to the approximate length required for the electrode;
\item Each piece is bent into a rectangular ``U''-shape by forcing it around the corners of a 13 mm ($\approx\pi\cdot$4 mm, found by trial and error) hardened steel plate;
\item In order to obtain the inner circular shape, each electrode is bent around a 4 mm stainless steel rod.
\end{enumerate}
Tantalum wire is used because of its high resilience to electrical discharges, and because it can be bent into the desired shape quite easily.\cite{Marian:2010kc}
The stretching of the wire is necessary to ensure straight wires, but it also reduces the diameter to 0.57 mm.
Step 2 is crucial because it ensures that the wires are bent at right angles at the point where the two ends of the circular bend meet.
All attempts to bend the electrodes directly into this shape, without the pre-bending, resulted in electrodes with cross sections that resembled teardrops rather than circles.
Considerable deviation from the overall cylindrical symmetry of the decelerator is obtained if the individual electrodes are not as close to circular as possible.
The above procedure guarantees that all 320 electrodes are identical to within high accuracy, and that the electrodes are produced without being scratched, thus rendering them more resistant to discharges when high voltages are applied.
The electrodes are then cleaned in an ultrasonic bath and mounted on the main electrode rods.
Cleaning is an important step in order to remove traces of the polishing material from the surface of the electrodes, and this step significantly enhances the resilience of the electrodes towards electrical discharges.

Each of the mounting rods contains 40 holes for the electrodes, one of which is seen in figure \ref{fig:fig1}D, and two tapped holes for the mounting.
The ring electrodes are pushed into the holes and pre-aligned using a PVC block that fixes the distance between the rod surface and the electrode bend.
The Ta electrodes are then fixed to the steel rods by pushing one or two pieces of bent, 0.3 mm piano wire into the spaces left between the Ta wire and the walls of the hole.
This ensures that the electrodes do not slip out, but leaves the required flexibility for final alignment.
 At this stage, the electrodes are positioned to within $\approx$0.5 mm, insufficient for the present purposes.

The exact positioning of the electrodes on the individual rods is complicated by the fact that their positions have to be measured without mechanical contact, both in order to avoid damage on the surface of the electrodes, and because any contact might bend the wires.
A solution is obtained by pre-mounting the rod with the ceramic mounts and the bronze bar (in the same combination as in the final setup) on an aluminum block.
This block is then positioned on a digital microscope that is built by fixing a CCD camera equipped with a macro-zoom lens to the head of a mill.
The aluminum block, together with the pre-mounted electrode assembly, is accurately positioned on the moving table of the mill.
In this way the relative positions of individual electrodes on each rod can be measured and set with a limit given by the resolution of the CCD.
By placing the mount with the electrodes horizontal or vertical, it is possible to align and position the electrodes relative to the main rod and relative to each other with an estimated accuracy of $\pm$0.02 mm.

Finally, the 8 rods are mounted on the main cylinder and aligned to each other.
This last step is done by positioning one CCD camera above and one to the side of the decelerator, and a theodolite below it.
Using custom-made alignment tools, each rod is positioned relative to its neighboring rods and to the main axis of the decelerator.
The vertical alignment can only be measured at one particular point, but the very accurate positioning of the individual electrodes on the mounting rod ensures the same degree of alignment throughout the entire structure.

\begin{table}[htdp]
\begin{center}
\begin{tabular}{|l|l|}
\hline
Parameter & Precision \\ \hline
Electrode diameter & $\pm$0.02 mm \\
Distance between adjacent electrodes &$\pm$0.02 mm \\
Electrode center to mounting rod & $\pm$0.02 mm\\
Electrode offset from beam axis &$\pm$0.02 mm\\
Electrode tilt around handle & $\pm$10 mrad\\
Rod in plane perpendicular to beam axis & $\pm$0.03 mm\\
Rod position along beam axis & $\pm$0.02 mm\\
Tilt of rod perpendicular to beam axis &$\pm$2 mrad\\ \hline
\end{tabular}
\caption{List of estimated precision for critical parameters of the ring electrode decelerator.}
\label{table:precision}
\end{center}
\end{table}%

A quantitative assessment of the overall precision in the alignment is difficult.
Table \ref{table:precision} summarizes the main estimated deviations from the ideal structure.
These estimates base on a critical evaluation of each of the production steps, as well as on test measurements on selected electrodes and electrode positions after the alignment.
The electrode diameter is defined through the bending around the 4 mm rod, and the deviation stated in the table has to be read as possible deviations from a perfectly round shape, as well as a deviation from the ideal diameter.
Due to the use of optical tools in the alignment procedure, the most difficult, and thus also least precise, part is the accurate positioning of the main electrode mounting rods in the plane perpendicular to the guide axis, mainly because of the mutual visual obstruction of the electrodes.

An attempt to characterize the effects of these deviations has been made by simulating the resulting electric fields in three dimensions using finite element calculations.\cite{comsol}
A horizontal displacement of a single electrode moves the trap off-center when it is located in the plane of that electrode and has little effect otherwise.
Displacement of an entire rod thus leads to a periodic jumping of the traps, with a frequency given by the modulation frequency.
Vertical displacement changes the electric field between two electrodes but has little effect on the field inside of the decelerator.
Longitudinal displacement of an entire rod can thus lead to a periodic deformation of the traps to produce harder or softer walls with a periodicity also determined by the modulation frequency.
While the modulation frequencies in the experiments done so far were substantially higher than the oscillation frequencies of the molecules inside the traps, which are in the range of 1 kHz, it still has to be assumed that the mechanical inaccuracies to some extent contribute to the increased losses observed experimentally when compared with the theoretical predictions.
In the case of longitudinal displacement, a higher risk is however the increased danger of electrical discharges.

\subsection{High voltage-waveform generation}\label{ch:el}
\begin{figure}
\includegraphics[width=\linewidth]{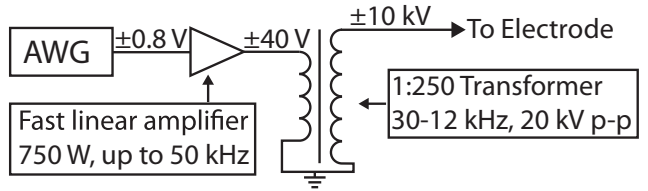}%
\caption{\label{fig:fig2}Simplified circuit diagram for the generation of the high-voltage waveforms. AWG designates the arbitrary waveform generator. Each of the eight channels in the current setup was driven by a separate amplification line like this.}
 \end{figure}

 A diagram of the electronics used to drive a single electrode rod is shown in figure \ref{fig:fig2}.
Computer-controlled waveforms with amplitudes of up to $\pm$1 V are initially generated in an arbitrary waveform generator (AWG; Wuntronic, DA8150) for all eight channels simultaneously.
Programming of the AWG gives full control over the phase and amplitude of each of the waveforms at every moment during the deceleration, and for each of the channels separately.
The AWG provides 150 MS/s at 12 bit vertical resolution which corresponds to $\approx$6 kS per period at 25 kHz modulation.
The 0.5 mV vertical resolution corresponds to only 5 V in the final waveform.
The amplification procedure filters out high frequencies, and limited vertical resolution produces a slight uncertainty in the final voltages applied to the electrodes.
However, considering the magnitude of other error sources, this effect can be neglected.
For every channel, there is a separate amplifier (Servowatt DCP 780/60 HSR) that raises the waveform amplitude from $\pm$0.8 V to $\pm$40 V.
The high power of each amplifier (750 W) is required to drive a high current ($\pm$25 A) through the primary winding of a transformer (custom made by \emph{Weiers und Partner}) to produce the final $\pm$10 kV waveform.
The phase of the individual channels is conserved during the amplification.
A slight frequency dependence of the overall-amplification is, however, observed.
In order to avoid excessive fluctuation of the trap shape, the amplitude on each electrode mounting rod is first checked as a function of frequency.
The observed deviations from a constant amplitude are then corrected by modifying the input waveform.
Since neighboring electrodes capacitively influence each other, this procedure has to be done iteratively to ensure constant and stable amplitudes.
The process converges fast, however, and not more than two iterations were necessary in general to produce waveform amplitudes constant to within a few percent.

\section{Results}\label{ch:sim}
The operation of the apparatus is demonstrated in the following sections.
In the first section, the phosphorescence of the metastable CO molecules is used to image their motion along the decelerator axis and the formation of density gradients near the trap positions.
In the second section, the transverse distributions are investigated with application of dc voltages to the electrodes, thus forming an electrostatic guide.
The third section describes the results of the imaging of molecules that were guided in modulated electric fields.
Finally the effect of potential non-adiabatic transitions in CO is studied by comparing arrival time distributions of \CO and \COtn.
These results complement the measurements presented previously.\cite{Osterwalder:2010bx}
There, arrival time distributions for \COt were presented for guiding with modulated potentials and for deceleration.

\subsection{Phosphorescence images of molecules in the decelerator}
\begin{figure}
\includegraphics[width=\linewidth]{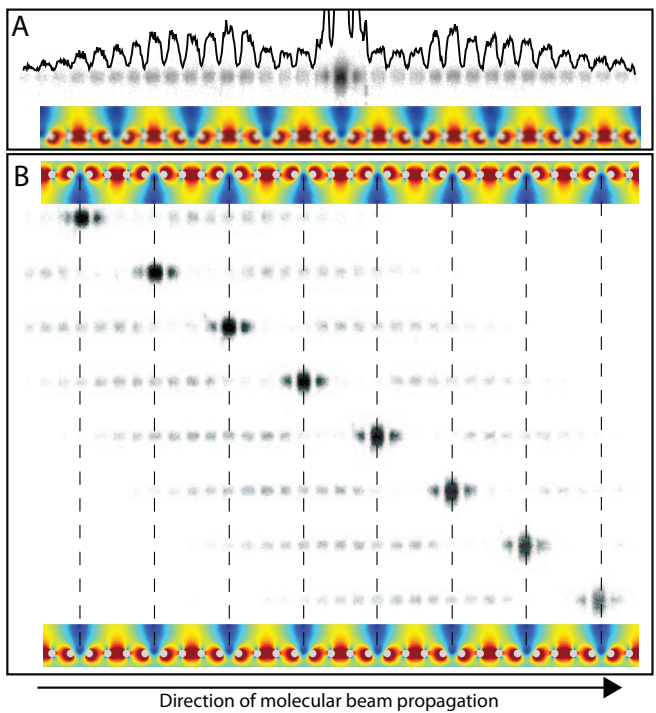}%
\caption{Phosphorescence images of CO packets flying through the decelerator structure, recorded with a CCD camera from the side of the decelerator chamber. The central part of panel A shows an image recorded with a delay of 617~\textmu s from the laser pulse and a width of 2~\textmu s. The delay is chosen to capture the trap exactly between two electrodes. The lower part of panel A shows the electric field magnitude at that time, and the curve at the top is a vertical integration of the image, showing the spatial dependence of the molecule density. The intensity scale is cut at $\approx$20\% of the peak maximum to emphasize the smaller peaks on the side (see text). The zeroes in the curve are the shadows of the electrodes. Panel B shows a series of images, starting at a delay of 525.5~\textmu s after the laser and with 5~\textmu s wide gates. The time delay between subsequent images is 20~\textmu s, corresponding to 1/2 period at 25 kHz. Above and below the images are plots of the electric field magnitude which has the same shape for each image. Vertical dashed lines show the calculated positions of the traps to guide the eye.\label{fig:fig4}}
 \end{figure}
The metastable state of CO used here has a lifetime of 2.63~ms,\cite{Gilijamse:2007wda} and it decays to the electronic ground state by emission of a photon with wavelength between 206 nm and 240 nm.\cite{Jongma:1997cm}
This phosphorescence can be imaged, using an image-intensified CCD camera, to follow the density of CO molecules as a function of time and position.

Figure \ref{fig:fig4} shows images taken from the side of the decelerator about 15~cm from its entrance (see figure \ref{fig:fig1}A).
These images were taken through the gaps between the decelerator electrodes, and the shadows of the electrodes are visible as a periodic intensity modulation.
The molecules were guided through the decelerator at a constant velocity of 300 m/s.
Figure \ref{fig:fig4}A shows a single 2~\textmu s-long snapshot recorded at a fixed delay of 617~\textmu s relative to the laser pulse that produces the metastable CO molecules.
This delay was chosen such that the densest part of the distribution would be located exactly between two electrodes.
The horizontal axis corresponds to the molecular beam axis, with the source on the left and the detector on the right.
The graph above the image shows the same data integrated over the transverse coordinate.
The intense region at the center of the image is produced by the single trap that is filled by the molecular beam; the false color image of the electric field magnitude at the bottom of the panel shows the position of the traps (darker areas) at the time of the snap-shot. 
On either side of the main peak are molecules that jumped over the barrier to the previous or next trap.
Because of their energy, they continue to pass over all saddle points along the decelerator.
Since they slow down on the saddle points, there is a slight increase in the density which is visible 6 electrode spacings away on either side of the main peak.
The structure becomes more pronounced by the time the molecules have passed through the entire decelerator, and they are very clear in the arrival time distributions shown, e.g., in figure \ref{fig:fig34}.
The detection efficiency of the imaging is relatively low.  
The 5 cm-diameter lens 35~cm from the center of the decelerator only collects 0.13\% of the light emitted by the phosphorescing molecules, and the imaging system itself is only about 10\% efficient.  
Additionally, during a 2~\textmu s exposure, only about 0.07\% of the molecules will phosphoresce.  
The resulting, overall detection efficiency is roughly $10^{-7}$.  
From this and the total number of photons observed in the images, we can infer that about 10$^7$ molecules are guided in the main trap at a density of approximately 10$^9$ molecules/cm$^3$.
%we can infer that several million molecules are guided in the main trap. 

By changing the time at which the image is taken, the motion of the cloud of molecules through the decelerator can be followed.  
Panel B of figure \ref{fig:fig4} shows a series of 5~\textmu s-wide images taken at 20~\textmu s intervals, starting at 525.5~\textmu s.
A guiding velocity of 300~m/s requires 25 kHz waveforms, and the time interval corresponds exactly to one half period.
Since there are two field minima per period, the electric field distribution (shown at the top and bottom of the panel) is the same for each image.
The dashed vertical lines show the positions of the traps at the respective times.
The images show a cloud of molecules that is confined both transversely and longitudinally as it is guided through the device in a single moving trap.
The molecules outside this trap are transversely confined, but will spread out longitudinally as they proceed through the decelerator; the only remaining structure in the arrival time distribution is given by the bunches formed on the saddle points, as described above.

\subsection{Guiding of molecules with static fields}\label{dcguiding}
\begin{figure}
\includegraphics{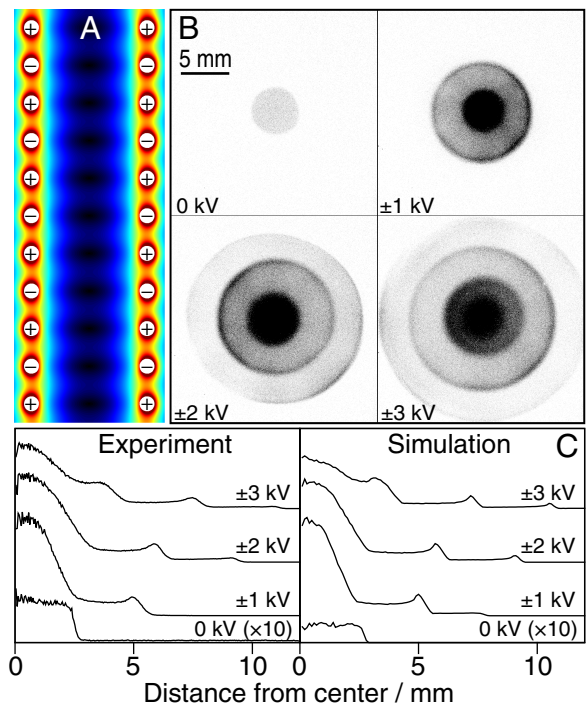}%
\caption{\label{fig:static}Transverse position distribution of the molecules after the decelerator with 0 kV, $\pm$1~kV, $\pm$2~kV, and $\pm$3~kV applied to the electrodes.  Panel A: electric field distribution in the decelerator with alternating positive and negative voltages on the electrodes. Panel B: two-dimensional transverse distributions at the detector for the various applied potentials. Panel C: angle-integrated radial intensity profile of the distributions shown in panel B.}
 \end{figure}
Errors in the operation of the decelerator can stem from imperfections in the construction and in the waveforms.
By applying static potentials to the electrodes, the errors due to imperfections in the construction can be examined independent of the waveform errors.
If alternating positive and negative potentials are applied to the electrodes, the resulting distribution of electric field magnitude has a minimum on axis (shown in panel A of figure \ref{fig:static}) and increases steeply near the walls of the tube.
This electric field distribution generates a guide that confines the molecules in low field seeking states transversely only.

Although there is also a longitudinal modulation of the electric field strength (including a point of zero electric field at the center of each electrode), this modulation is too weak to significantly affect the longitudinal motion of the molecules; application of guiding potentials increases the flux without significantly changing the arrival time distribution. 
The transverse position distribution several centimeters behind the decelerator, on the other hand, shows significant structure that can be used to characterize the decelerator.
To measure this distribution, an imaging detector (microchannel plate (MCP) detector, coupled to a phosphor screen) is placed 18~cm behind the last electrode of the decelerator.
Metastable CO molecules that hit the MCP generate an Auger electron which is subsequently amplified by the MCP stack.
The resulting spot on the phosphor screen is recorded on a CCD camera.  
The center position of each spot is determined using a centroiding algorithm. 
By switching the voltage applied to the front of the MCP stack, detection is restricted to molecules with an arrival time corresponding to a longitudinal velocity of 300$\pm$6 m/s.

The images in figure \ref{fig:static}B show the position distributions of molecules at the MCP with 0~kV, $\pm$1~kV, $\pm$2~kV, or $\pm$3~kV applied to the electrodes of the decelerator.  
At zero volts (upper left image in figure \ref{fig:static}B), the molecules form a circular distribution, the size of which is mainly determined by the geometric aperture formed by the diameter and length of the guide, and the distance to the imaging detector.
Two main changes are observed as the potential on the electrodes increases: 1. The overall intensity increases, and 2. in addition to the intense central disk, larger rings appear that grow with increasing voltage.
As the potential is increased, molecules with larger transverse velocities can be guided through, resulting in the increased overall intensity.
These molecules also execute more oscillations in the device, producing more rings that grow in diameter.
While the images are almost circularly symmetric, some asymmetries are also visible.
The prominent left-right asymmetry, particularly in the outer rings, results in part from an off-center placement of the laser excitation at the beginning of the experiment, and it can be influenced by moving the laser.
An up-down asymmetry of a similar degree is obtained by slightly detuning the laser.
This detuning leads to the selective excitation of molecules with a velocity component along the laser beam that are Doppler-shifted into resonance.
The bandwidth of the laser corresponds to a velocity range of $\approx$30 m/s which is in a range similar to the acceptance of the guide.
A subtle eight-fold symmetry is also visible in some of the rings, which results from the eight different positions of the electrode indentations. 
These result from the shape that was chosen for the ring electrodes here (see figure \ref{fig:fig1}D).
The subtlety of this structure in the images indicates that the indentations do not significantly affect the operation of the decelerator.

The left (right) panel of figure \ref{fig:static}C shows the experimental (calculated) angle-integrated radial intensity profile of the images in panel B.
The calculations were done via two-dimensional trajectory simulations that assume cylindrical symmetry.  
In all cases except with no voltage applied, the relative number of molecules reaching the detector is accurately reproduced by the trajectory simulations.
In the field free case, the lower number of molecules reaching the detector in the simulations can be qualitatively explained by the existence of non-low field seeking states; these states are present in the experiment and travel along the same straight-line trajectories as the molecules in the low field seeking state when no electric fields are applied, but are not included in the trajectory simulations.
The ring diameters are accurately reproduced by the simulations, but the simulated rings have sharper edges than are observed experimentally.
Since the peaks at the ring radii are wider in the experimental data than in the simulated images, the contrast is also reduced.
As a consequence some of the rings, in particular the outermost ones, are barely visible in the experimental images while they are still clearly visible in the simulations.
Simulations show that the rings stem from molecules that pass close to the electrodes at the end of the decelerator, and that are slightly focused.
The intensity distribution in these rings depends sensitively on the electric field distribution just behind the decelerator.  
This region is more strongly influenced by the support structure of the electrodes, which is not included in the cylindrically-symmetric model.

Overall, the measurements show only minor deviations from the predictions of the two-dimensional trajectory simulations, indicating that the misalignments of the decelerator are relatively small.
Remaining differences are attributed mostly to non-cylindrically symmetric fringe-fields behind the decelerator.
These distort the transverse velocity distribution of the molecules emerging from the decelerator by asymmetric beam-focusing or -defocusing.
They have not been included in the present calculations, and to do so would indeed be quite demanding, both because the fields would have to be accurately known, and because the calculations would have to be done in three dimensions, without assuming cylindrical symmetry.

\subsection{Guiding of molecules with AC fields}
\begin{figure}
\includegraphics{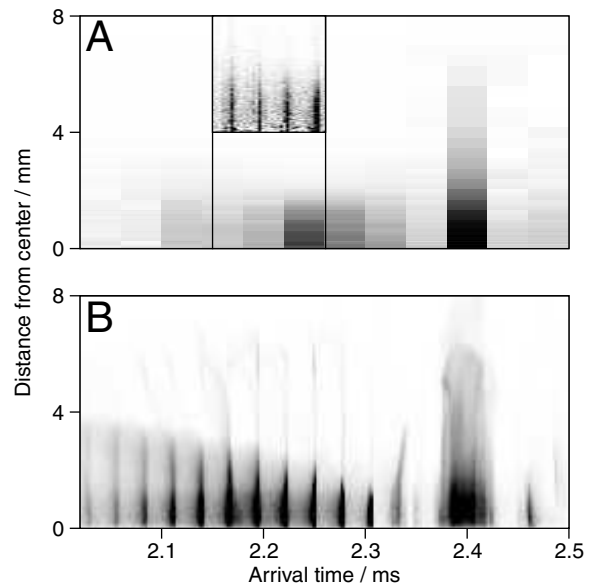}%
\caption{\label{fig:img}Panel A: Angle-integrated radial intensity profiles of the two-dimensional transverse distribution at the detector after actively guiding molecules at 288 m/s.  The inset shows the narrow region below it with a higher time resolution.  Panel B: Results from trajectory simulations under the same conditions.}
 \end{figure}
Similar images to those shown in the previous section were also recorded after actively guiding the molecules at 288 m/s.
Panel A of figure \ref{fig:img} shows the angle-integrated radial intensity profiles of the transverse distribution for the range of arrival times covering the fast part of an initial distribution centered around $\approx$300 m/s.
The main panel shows the distributions measured in 40~\textmu s-wide arrival time gates over a large range.
Such slices are obtained by applying a short voltage-pulse to the front of the MCP-stack in the imaging detector.
The fast part of the initial molecular beam, which arrives around 2.2--2.3 ms, shows a narrow transverse distribution; these molecules move faster than the traps and are not confined by the longitudinal potential.
They traverse the saddle points between traps and thus see a much weaker transverse confining potential.
The inset of panel A shows the narrow range of arrival times below it with a higher resolution (3~\textmu s wide gates).
The narrow peaks that result as the molecules are bunched while traversing the saddle points are visible here.
The longitudinally-confined guided molecules are visible at an arrival time of 2.4~ms.
Because these molecules are confined in a deep potential well, a much wider transverse velocity distribution can be guided to the end of the decelerator.
Panel B shows the results of numerical trajectory simulations that agree well with the experimental data.

\subsection{Non-adiabatic losses}
\begin{figure}
\includegraphics{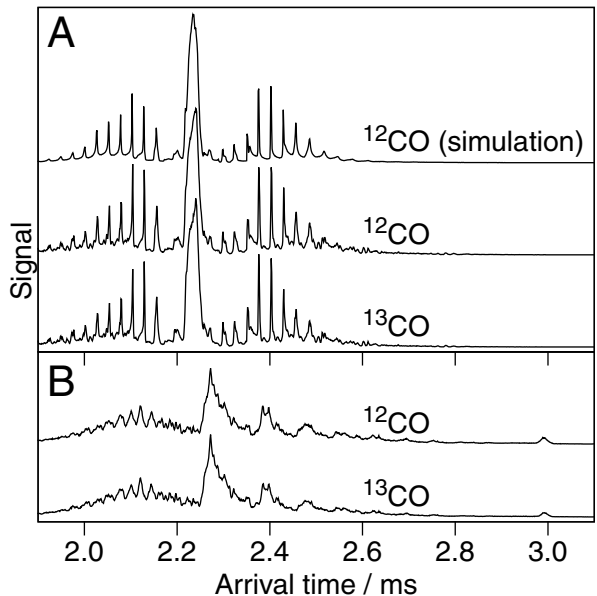}%
\caption{\label{fig:fig34}Comparison of the arrival time distributions for \CO and \COt for guiding at 300~m/s (panel A) and deceleration from 300~m/s to 180~m/s in 2~ms (panel B).}
 \end{figure}
In the Stark decelerator on a chip, it was observed that non-adiabatic transitions from the low field seeking states to states only weakly affected by the electric fields result in significant losses of molecules from the traps.
Similar losses have been observed for ammonia molecules trapped in macroscopic electrostatic traps, albeit on a longer time scale.\cite{Kirste:2009cf}
The chip decelerator operates on a similar principle as the one described here, in that the required fields for the deceleration are obtained via the application of sine-modulated potentials to a periodic electrode-array.
It has been shown that, in order to decelerate molecules on this chip, the non-adiabatic losses must be prevented by either using the \COt isotopologue\cite{Meek:2009er} or by applying an external magnetic field. \cite{Meek:2011gs}
Non-adiabatic losses of molecules from the traps on the chip result primarily from the small size of the traps: as the molecules pass near the zero-field trap center, the direction of the electric field vector changes rapidly, providing spectral components that can drive the transitions to non-trapped states.  
In the macroscopic traps of the decelerator presented here the rate of such transitions should be much lower.

To confirm that the effect of such losses is small, $^{12}$C$^{16}$O and $^{13}$C$^{16}$O, were both guided at 300 m/s and decelerated from 300 m/s to 180 m/s over the length of the decelerator, and the arrival time distribution of the molecules exiting the decelerator was recorded.   
Because of the hyperfine structure in \COtn, any losses due to non-adiabatic transitions should be reduced in comparison with \COn.\cite{Meek:2009er}

Figure \ref{fig:fig34}A shows, from bottom to top, the experimental arrival time distribution for \COt and \COn, and the simulation for \CO when guiding of the molecules at 300 m/s.
In all three traces, the main peak shows the molecules that are moving in the main trap, at 300 m/s, and the sharp and regular structures to either side of the main peak are from molecules that are bunched as they cross the saddle points between the traps.
Except for an overall intensity difference, the three traces show no significant differences.
The source of the different intensities lies mainly in different source conditions.
Additional information on potential non-adiabatic losses can be obtained by comparing the traces for deceleration: this is shown in \ref{fig:fig34}B for \COt (lower trace) and \CO (upper trace).
By normalizing these traces based on the relative intensities of the guiding data, it is found that about 60\% more \CO is decelerated than \COtn.
This difference in deceleration efficiency is qualitatively explained by the slightly lower mass of \COn.  
Significant non-adiabatic losses would lead to an enhanced efficiency of \COt deceleration.

Monte-Carlo trajectory simulations for guiding of \CO (shown as uppermost trace of figure \ref{fig:fig34}A) and \COt (not shown) show excellent agreement with the experimental data.
The relative amounts of \CO and \COt that can be decelerated also agree well with the predictions of the simulations, which do not allow for non-adiabatic processes.  
On the timescale of the current experiments (which is similar to the radiative lifetime of the metastable molecular state that is used), losses due to non-adiabatic transitions are thus concluded to be negligible for the case of CO.

\section{Summary and Outlook}\label{ch:sum}
A detailed account on the design, construction, and operation of the recently developed traveling wave decelerator for polar neutral molecules has been given.
The very high accuracy required for the mechanical design calls for special care during the production and alignment of the individual electrodes.
Generating high-voltage sinusoidal waveforms with frequencies between 25~kHz (for the current geometry) and ideally 0 Hz is particularly challenging.
In the currently employed design, the range 12-26 kHz is accessible, and this could be extended by using different transformers.
To decelerate molecules from higher initial velocities or to lower final velocities one could segment the decelerator into different parts that are each driven by a separate set of transformers.
By having slightly overlapping frequency ranges between neighboring segments, a smooth deceleration would then be possible over wider velocity ranges, albeit at the cost of flexibility.
A more flexible solution, that however requires more development, will be the use of a linear amplifier that covers the entire range of interest.

The good agreement between the simulations and experiments raises the confidence in the understanding of the overall dynamics of molecules inside this type of decelerator.
The longitudinal distribution is measured by recording the molecule arrival times, and the transverse distribution has been characterized via imaging of the complete packet of guided molecules in dc-electric fields.
Narrow time-slices of the transverse distributions have been recorded for guiding in modulated fields, and it was found that the calculated and measured data agree reasonably well.
Remaining discrepancies, for dc- and ac-guided molecules, are attributed to fringe fields that are generated from the electrode mounting structure.
These fields are shielded from molecules that are inside the decelerator, but not from molecules flying from the end of the decelerator to the detector.
The comparison between \CO and \COt showed that in the present geometry, unlike the case of the chip decelerator, non-adiabatic losses can be completely neglected.
This is because the increased trap-size leads to lower additional frequencies perceived by the molecules as they fly through the central region of the traps.

As has been suggested previously,\cite{Osterwalder:2010bx} this type decelerator seems particularly suitable for the deceleration of heavy diatomic molecules.
Each rotational state of these molecules has low-field-seeking components only below a certain electric field.
Above that, all levels undergo an avoided crossing with levels from the next rotational state and become high-field-seekers, making them inaccessible by traditional Stark deceleration.
The only option so far was the use of alternating-gradient deceleration, but this approach has not yet been demonstrated to work stably and efficiently.
In the present decelerator, however, the electric field magnitude inside the moving traps is always relatively low, yet the high electric field gradient nevertheless enables efficient deceleration.

\section{Acknowledgments}
This work is funded through the Max-Planck-Society, the Ecole Polytechnique F\'ed\'erale de Lausanne, the Swiss National Science Foundation (grant number P002-119081), and through ERC-2009-AdG under grant agreement 247142-MolChip.
%\bibliographystyle{aip}
%\bibliography{macro}

\end{document}